\documentclass[openacc]{rsproca_new}

\begin{document}

\title{Research on the identification of the two-phase flow pattern of gas-liquid in a vertical rising tube based on BP neural networks }

\author{
	Xiaojun.Zhang$^{1}$, Shijiao.Liu$^{2}$ ,Jiayue.Qian$^{3}$,Xingpeng.Shen $^{4}$and Jianlong.Liu$^{5}$}

\address{$^{1,2,3,4,5}$College of Physics and Optoelectronic Engineering, Harbin Engineering University, Harbin, China}





\begin{abstract}
Research on the identification of the two-phase flow pattern of gas-liquid in a vertical rising pipe is of great significance for improving the production capacity and production efficiency of the petrochemical industry. In order to address the problem of the accuracy of the identification of the two-phase flow pattern of gas-liquid, this paper proposes a method for identifying the two-phase flow pattern of gas-liquid in a vertical rising pipe based on BP neural networks. In the study, the Fluent software was used to numerically simulate different two-phase flow velocities. The pipes were all constructed as vertical rising pipes with an inner diameter of 20 mm and a length of 2000 mm. Three flow pattern cloud diagrams and their related data were obtained for bubble flow, elastic flow, and annular flow. The gas content of the three flow types was used to collect data to form a database. The BP neural network was used to classify and identify the three flow patterns, but the result was only 90.73$\%$. We again used the Adam algorithm to optimise the BP neural network and regularise it, and the flow pattern recognition result reached 96.68$\%$, which was a better recognition 
\end{abstract}




\maketitle

\section{introduction}
As a common and complex flow phenomenon [1-2], two-phase flow is widely found in many engineering fields, such as petroleum engineering, chemical engineering, nuclear engineering, etc [3-5], and plays a vital role. In these fields, the behaviour of two-phase flow directly affects the efficiency and safety of the process. Therefore, it is particularly important to study its flow characteristics in depth. Among them, a vertical rising pipe is a common device for two-phase flow, which is widely used in various engineering systems.

In a vertical rising pipe, five main two-phase flow patterns can be formed: bubble flow, elastic flow, stirred flow, annular flow and annular jet flow. These flow patterns have their own characteristics and their hydrodynamic properties differ significantly. For example, bubble flow is mainly characterised by bubbles dispersed in the liquid and is suitable for low flow rates. In contrast, annular jet flow is characterised by strong interaction between the gas and liquid at higher flow rates [6]. The flow characteristics of different flow patterns affect the pressure loss, heat transfer efficiency and mixing effect of the fluid [7-8]. Therefore, a thorough understanding and control of these flow patterns is crucial for the design and optimisation of related processes. Bubble, elastic and annular flows are the three most common flow patterns, so this paper focuses on identifying these three flow patterns.

However, traditional two-phase flow pattern prediction methods often rely on experience and experimentation, which is not only inefficient, but also difficult to adapt to the needs of increasingly complex working conditions. With the continuous advancement of computer technology and numerical simulation methods, researchers can identify two-phase flow patterns based on parameter characteristics. They can also identify the formation mechanisms of different flow patterns and their transition conditions through careful analysis of massive amounts of data [9, 10]. At present, there are electrical tomography, differential pressure method, annular conductivity method, optical fiber probe method, and ultrasonic method for measuring two-phase flow parameters [11, 12, 13, 14, 15]. The identification of two-phase flow patterns is carried out using data-driven methods, which can significantly improve the efficiency and accuracy of two-phase flow pattern recognition [16]. For example, Zhang Lifeng et al. proposed a two-phase flow pattern recognition method for gas-liquid in vertical pipes based on convolutional neural networks (CNNs) and gated recurrent units (GRUs), using reconstructed images from an electrical resistance tomography (ERT) system. The recognition rate reached 99$\%$ [17]. Shi Yanyan et al. proposed a gas-liquid two-phase flow pattern recognition method based on multi-feature fusion and grey wolf optimization support vector machine (GWO-SVM), with an identification rate of 98.45$\%$ [18]. These two-phase flow pattern recognition methods not only improve the intuitiveness of the research, but also provide a more reliable basis for practical applications. However, the current two-phase flow pattern recognition algorithm relies on detection technology, which is time-consuming and expensive in the complex and ever-changing downhole environment [19], so it has not been widely promoted. In view of this challenge, it is urgent to develop other detection methods and algorithms to ensure that the required data can be accurately obtained under complex working conditions and provide a reliable basis for decision-making.

In recent years, some scholars have used Fluent software to simulate two-phase flow patterns and compare the simulation results with experimental results, and the simulation results are reasonable [20]. In order to accurately predict the change of flow pattern, we will rely on computational fluid dynamics software Fluent to simulate the void fraction measurement results of two-phase flow patterns measured by fiber optic probe method, establish a high-precision neural network model using BPNN, systematically train and verify it, and continuously optimize the algorithm to improve the flow pattern recognition rate. Through this research, we hope to provide a simple and efficient method to predict the flow pattern of gas-liquid two-phase flow in a vertical rising tube, so as to provide a reliable basis for the design and optimisation of related engineering fields. The successful implementation of this method will open up new perspectives for the research and application of gas-liquid two-phase flow and promote the development and progress of related technologies.

\section{ Numerical simulation of flow patterns}

The object of this paper is a vertical rising pipe with a radius of 10 mm and a height of 2000 mm. The aim is to explore the characteristics of different flow patterns through experiments on gas-liquid two-phase flow. During the experiment, we will obtain a variety of flow patterns by adjusting the different flow rates of the gas phase and liquid phase, and analyze their gas content distribution. To achieve this goal, the experiment is divided into the following steps:

\subsection{Experimental model construction}
First, this paper will collect a large number of result parameters from numerical simulations to provide a reliable data basis for training and verifying the neural network model. The experimental data will be numerically simulated using the computational fluid dynamics software FLUENT, focusing on the flow behaviour of different flow patterns in a vertical rising pipe, and recording the void fraction at different coordinates of each model.
During the simulation process, a 2D geometric model was constructed in the simulation software based on the actual device diagram. As shown in Figure 1, Figure 1 is a vertical rising pipe rotated 90, and Figure 1(a) is a geometric model of a vertical rising pipe. The bottom of the geometric model is set as the inlet of two-phase flow, while the top of the vertical rising pipe is the outlet of gas-liquid two-phase flow, and the two sides are the walls. The pipe is designed to be rectangular in order to facilitate the analysis of fluid flow. Next, the geometry was meshed as shown in Figure 1(b). In order to accurately capture the flow characteristics of the two-phase flow in the pipe and take into account the influence of the pipe wall on the flow, the pipe wall was doubled. After a mesh independence verification study, the final number of meshes selected was 122,520. This fine mesh will provide more accurate results for subsequent flow simulations.

 \begin{figure}[!h]
 \centering
 \includegraphics
   [width=0.65\hsize]
   {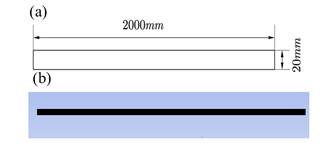}
   \vspace{-2 em}
 \caption{ Figure 1.Vertical rising pipe geometry and meshing. (a) Geometry (b) Meshing. The vertical rising pipe is designed to be 20 mm in diameter and 2000 mm high, with a two-phase flow inlet on the lower side and a two-phase flow outlet on the upper side. The left and right sides are walls. Figure (a) shows the geometric model of the designed pipe, and Figure (b) shows the meshing of the designed vertical rising pipe. The meshes on both sides of the wall are densified. In order to make the calculation more accurate and save time, the final number of meshes designed is 122,520. }
 \label{fig:one-column-figure}
 \end{figure} 
    \vspace{-2 em}

 In this paper, a VOF model is used coupled with a Levelset function, and the Pressure solver is selected for transient simulation. Considering that in a real experiment, the two-phase flow in a vertical rising pipe is affected by gravity, we set the gravity in the vertical direction to -9.81 and checked the box for implicit body force. The fluid phase is set to two-phase, with the materials being liquid water and gaseous air respectively. At the same time, the surface tension coefficients of air and liquid water are set to 0.072 N/m, and the standard two-equation turbulence model is used. The momentum equation is considered during the calculation, and the equation is as follows:

\begin{equation}
\frac{\partial \vec{u}}{\partial t}+E^{c}=E^{v}+S
\end{equation}

where: $ E^{c} $ is the convective term;  $E^{v} $ is the diffusive term; and  S  is the source term. The pipe inlet is a mixing velocity inlet.

\subsection{Data collection and pre-processing}
To facilitate the acquisition of flow data, we treat liquid water and air as incompressible ideal fluids and ignore their heat transfer characteristics to simplify the model. The data was collected as follows: we collected the gas content along a line parallel to the y-axis, and collected 17 air volume fractions on each line to ensure that the complete structural characteristics of the flow pattern were captured. By experimenting with different gas-liquid flow rate settings, we obtained a total of 504 sets of data, with 168 sets of data for each flow pattern.

We subjected the collected experimental data results to a series of preprocessing steps to ensure their suitability for training and verifying the neural network model. These preprocessing steps include data cleaning, data normalisation and feature extraction. First, data cleaning is performed to ensure data quality. We identify and remove duplicate, missing or abnormal data points to avoid interference with model training. Feature extraction is performed to convert the raw data into a form that is easier for the model to process. By selecting key features or generating new features, we can enhance the model's expressive power and improve its learning of complex patterns. After these preprocessing steps, the data will become more standardized and structured, providing a solid foundation for the input of the neural network model, thereby improving the accuracy and efficiency of predictions.
\subsection{ Characteristics of gas-liquid two-phase flow patterns}
(a) Bubble flow: As shown in Figure 2(a), the main characteristic of bubble flow is that the gas phase is discontinuous, with the gas distributed in the form of small bubbles in a continuous liquid flow. These bubbles usually have a spherical or elongated shape. In the middle of the pipe, the density of small bubbles is relatively high. When the gas phase content is small, the bubbles are small; as the gas content increases, some bubbles become larger, but they are still considered to be part of the bubble flow because the gas is not aggregated and the bubble size is much smaller than the pipe diameter. At this time, the gas flow rate is low, while the liquid flow rate is high. Therefore, the numerical characteristics of bubble flow are that there are more low flow pattern numbers and fewer high gas content numbers.
b) Bullet flow As shown in Figure 2 (b), bullet flow is characterised by alternating large bubbles and liquid masses, with the bubbles separated from the pipe wall by a liquid film. The bubbles are long, with the large bubbles taking the shape of bullets, i.e. Taylor bubbles, and the tails of the large bubbles in the flow often accompanied by many small bubbles. The numerical characteristics of the void fraction in bullet flow are that the values for high void fractions are clustered together, followed by values for lower void fractions. The air content cloud map of the bullet-shaped flow shows that the large bubbles are bullet-shaped because the air flow rate is high and the air is affected by the resistance of the water. From the perspective of flow pattern transformation, the flow rate of the bullet-shaped flow becomes larger, and the flow rate becomes larger. The increase in small bubbles causes them to aggregate into large bubbles, which are close in diameter to the pipe diameter. This flow pattern is a transitional zone between bubble flow and annular flow. The large bubbles are subject to water resistance and therefore take on a bullet shape. Since the large bubbles are the result of the aggregation and expansion of small bubbles, and the air flow rate is not particularly high, some small bubbles will not be able to complete the aggregation and expansion process. Therefore, in the cloud diagram, you can see that the tail of the large bubble is accompanied by many small bubbles. In practical applications, as the system pressure increases and the surface tension of the liquid decreases, the formation of large bubbles is restricted, so the range of existence of the elastic flow is relatively small. When the pressure exceeds 10 MPa, the elastic flow is often difficult to observe because it then transforms into a ring flow.
(c) Annular flow: As shown in Figure 2(c), annular flow is characterised by continuous liquid phase flow along the pipe wall, which is due to the viscous force of the liquid phase near the wall surface and the upward force of the liquid phase below. The centre is a continuous gas flow, and the interface of the air column is not parallel to the wall, which also poses certain problems for judging the flow pattern. This is due to the large gas phase flow rate and the small liquid phase flow rate; the gas phase flow rate is also large, so there are no small bubbles in the annular flow. There are also small liquid droplets in the air column, which is because there is a fluctuating interface between the liquid film and the gas-phase core flow. The fluctuations may cause the liquid film to break, allowing the droplets to enter the gas-phase core flow. Under certain conditions, droplets in the gas-phase core flow can also return to the liquid film on the wall. This flow pattern occupies the largest area among the two-phase flows and is one of the most typical flow patterns. Therefore, the numerical characteristics of the annular flow show that there are many values with high void fraction, while there are very few values with low void fraction.
   \vspace{-2 em}
  \begin{figure}[!h]
  \centering
  \includegraphics
    [width=0.65\hsize]
    {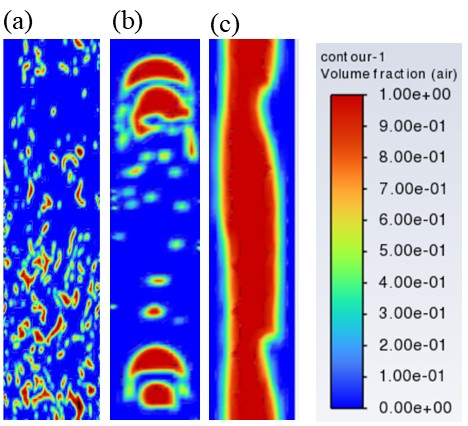}
  
  \caption{Figure 2 .shows a partial simulation of the gas-phase diagram for a vertical rising pipe with a two-phase flow pattern. The gas content is indicated by the different colours. (a) shows bubble flow, with bubbles that are either spherical or elongated. In the middle of the pipe, there is a relatively high density of small bubbles. (b) shows elastic flow, with large bubbles that are bullet-shaped, i.e. Taylor bubbles, located in the centre of the pipe and with a diameter close to the pipe diameter. The tails of the large bubbles in the flow are often accompanied by many small bubbles. (c) is annular flow, with a continuous core of gas in the centre of the pipe, which is interspersed with a small number of small droplets.}
  \label{fig:one-column-figure}
  \end{figure} 
   \vspace{-2 em}

The void fraction of the gas-liquid two-phase flow in a vertical rising circular pipe reveals the importance of the flow pattern characteristics and the flow pattern transition mechanism of the two-phase flow in the pipe. In this paper, 11 lines parallel to the y-axis were intercepted during the numerical simulation, with a spacing of 2 mm between each line, to obtain the void fraction at each point on the line, and the void fraction was averaged. The cross-sectional void fraction distributions of the three flow patterns are shown in Figure 3. Figure 3 shows the void fraction distribution of the three typical flow patterns in a vertical rising circular pipe. As can be seen from the figure, in the bubble flow, small bubbles are evenly distributed in the pipe. The void fraction in the centre of the pipe is slightly higher than that of the wall, and the cross-sectional void fraction is relatively stable, maintaining at about 0.24. The cross-sectional void fractions of the elastic and annular flows show significant differences: specifically, the void fraction near the centre of the pipe is significantly higher than that near the wall. In the case of the elastic flow, the Taylor bubbles are concentrated in the centre of the pipe and are almost the same size as the pipe, resulting in a void fraction of 0.40 in the centre, which is much higher than near the wall. Near the wall, due to the bubble's rolling suction effect on the liquid film, a small amount of bubbles are entrained in the liquid phase, and the void fraction drops significantly, eventually reaching a lower value. In the annular flow, the centre of the pipe is a continuous gas core with small liquid droplets entrained, and the cross-sectional void fraction is 0.81 and remains relatively stable. Because the liquid phase has a small flow rate and the gas phase has a large flow rate, the gas phase comes together throughout the two-phase flow process to form a gas core. There is a rolling suction effect between the gas core and the wall surface, and a thin liquid film is formed near the wall surface due to the viscous force of the pipe wall, which significantly reduces the void fraction near the wall surface. By analysing the cross-sectional void fraction of these three flow patterns, we can gain a deep understanding of the complex behaviour of gas-liquid two-phase flow in a vertically rising circular pipe [21]. As shown in Figure 3.
    \vspace{-1 em}
 \begin{figure}[!h]
	\centering
	\includegraphics
	[width=0.65\hsize]
	{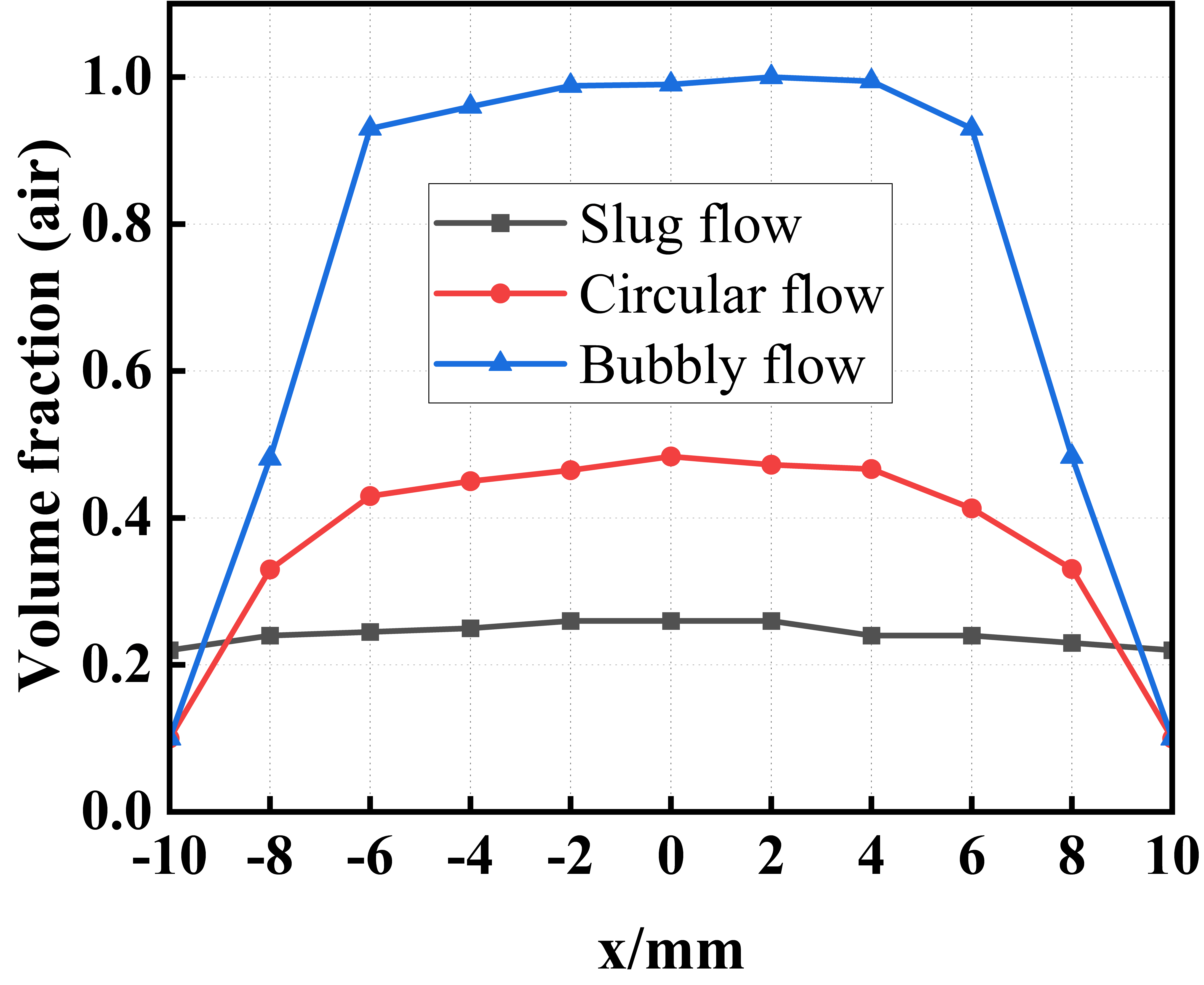}
	\caption{  Figure 3. Distribution of void fraction at different flow patterns in the vertical tube. The blue line represents the distribution of void fraction at the cross-section of annular flow, and the average void fraction is 0.40. The red line represents the distribution of void fraction of elastic flow, and the average void fraction is 0.81. The black line represents the distribution of void fraction of bubble flow, and the average void fraction is 0.24.}
	\label{fig:one-column-figure}
\end{figure}
   \vspace{-2 em} 
\section{Constructed and optimised BPNN}
The data collected is used to train and optimise a BP neural network [22] model. During training, input data is fed into the neural network, which generates an output result that is compared with the actual flow pattern to calculate the prediction error. Next, the error backpropagation algorithm is used to adjust the weights and biases in the network, thereby continuously improving the network's predictive ability. Training usually requires multiple iterations until the set training target is reached or convergence is achieved.
\subsection{ Constructed BPNN}

Based on the artificial intelligence-based BP neural network [23], we modelled the algorithm for identifying the flow pattern of gas-liquid two-phase flow in a vertical rising pipe. The process involves three steps: constructing, training and identifying the BP neural network. The flow chart of the BP neural network constructed in this paper is shown in Figure 4. Figure 4 shows a network structure consisting of three layers: an input layer, an output layer and a hidden layer. X represents the input signal, the number represents the hidden layer, and O represents the output layer, is the weight and b is the threshold parameter. The input layer has 17 neurons, and the output layer contains three categories: bubble flow, jet flow and loop flow. According to formula (2), the hidden layer is set to 25 neurons.

	 \begin{equation}
	 n_{l}=n+0.618(n-m)
	 \end{equation}

Where: n is the number of nodes in the input layer, m is the number of nodes in the output layer, and nl is the number of nodes in the hidden layer.
   \vspace{-2 em}
   \begin{figure}[!h]
    \centering
    \includegraphics
      [width=0.65\hsize]
      {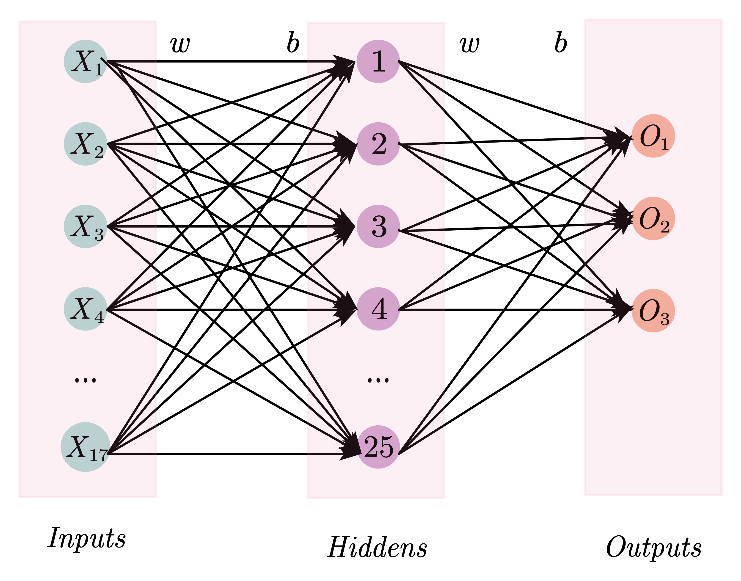}
    \caption{ Figure 4 .BP neural network structure diagram. The input layer has 17 nodes, the hidden layer has 25 nodes, and the output layer has 3 nodes.
    }
    \label{fig:one-column-figure}
    \end{figure} 
  
   \vspace{-2 em}
The constructed neural network was comprehensively trained and tested using 504 sets of data obtained from the experiment. We selected 353 sets of data as the training set and another 151 sets as the test set. The maximum number of iterations for the designed neural network was 100, the target training error was set to 10-6, and the learning rate was 0.01. For convenience, bubble flow was replaced by 1, spring flow by 2, and circulation by 3. The experimental results show that the BP neural network achieved the best validation performance in the fourth iteration, with an iteration error of 0.04. In the end, the test group achieved an accuracy of 90.73$\%$, but the algorithm can still be further optimised to obtain better results. The specific details are shown in Figure 5 below:

    \begin{figure}[!h]
     \centering
     \includegraphics
       [width=0.8\hsize]
       {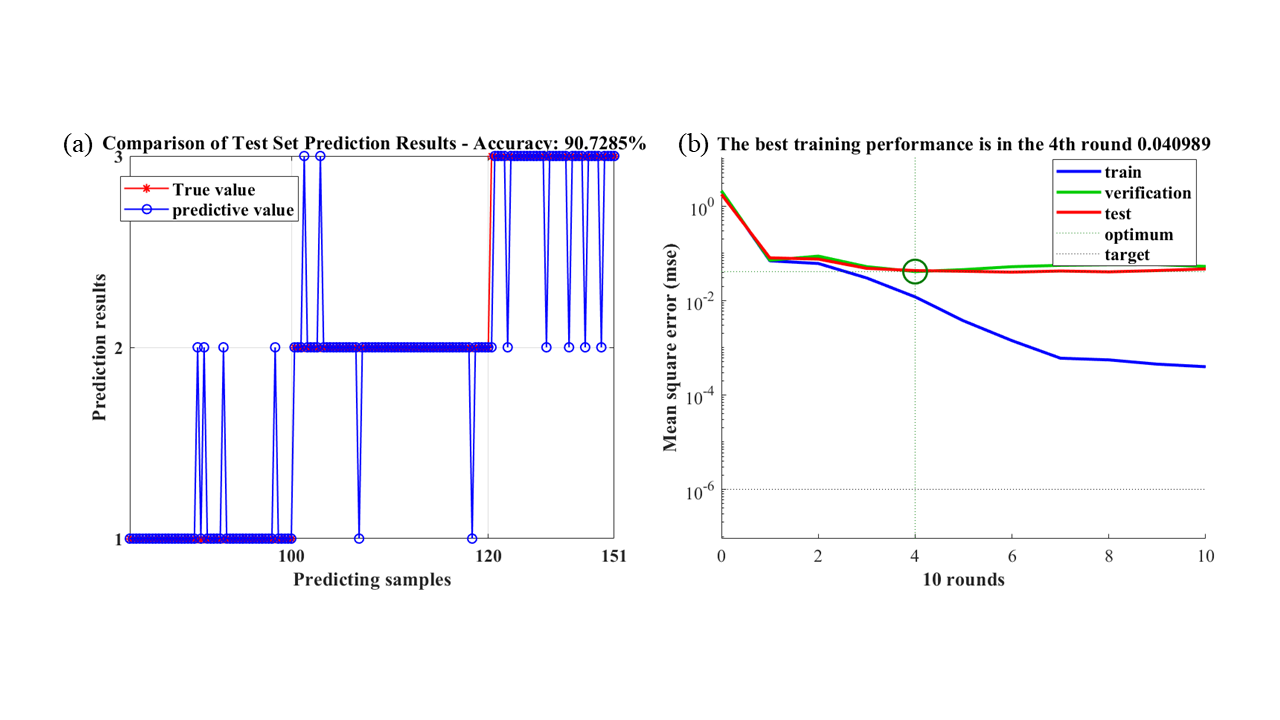}
 \vspace{-2 em}
     \caption{ Figure 5. BP neural network results. (a) shows that the prediction result accuracy of the test set is 90.73$\%$, and (b) is the mean square error of the first 10 iterations. The BP neural network achieved the best validation performance in the fourth round of iterations, when the iteration error was 0.04.
     }
     \label{fig:one-column-figure}
     \end{figure} 

\subsection{ optimised BPNN}
When using a BP neural network for flow pattern recognition, we aim to achieve an accurate fit of the output by adjusting parameters to perform various nonlinear transformations on the input data. To quickly find the optimal parameter solution and improve the flow pattern recognition rate, we chose to use the Adam optimizer [24] in combination with regularisation. When using the Adam optimizer, a regularisation term can usually be added to the loss function, so that the optimisation process takes into account both the prediction error of the model and the regularisation term. In this way, the Adam optimizer will apply regularisation when updating parameters, thereby improving the performance of the network, which can lead to better training results and higher generalisation capabilities. Figure 6 shows a schematic diagram of the optimisation process:

     \begin{figure}[!h]
      \centering
      \includegraphics
        [width=0.65\hsize]
        {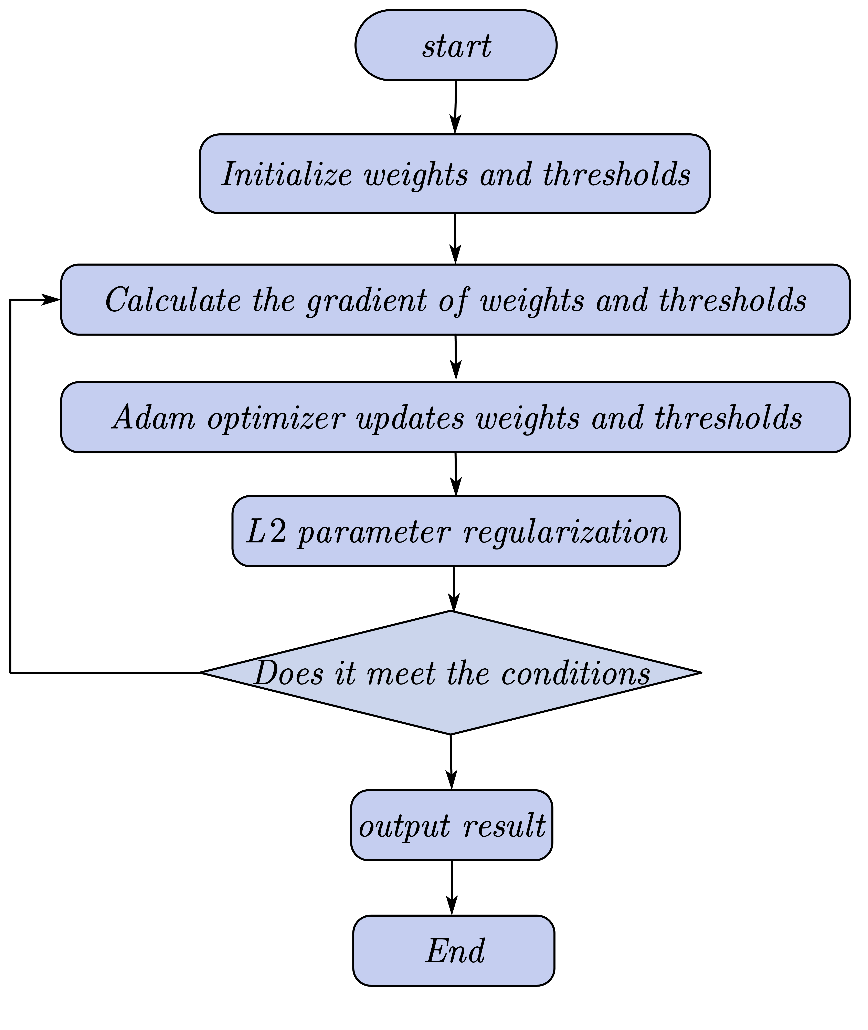}
           \vspace{-2 em}
      \caption{ Figure 6Regularization-Adam-BPNN flowchart. The designed BPNN uses the Adam algorithm and L2 regularization to optimize the algorithm to improve the flow pattern recognition rate
      }
      \label{fig:one-column-figure}
      \end{figure} 

\subsubsection{Theory and formula}

The key to the Adam algorithm is to simultaneously calculate the first moment (mean value) and second moment (uncentered variance) of the gradient and take an exponentially moving average of them, and to correct them for bias to ensure that the gradient estimate is not biased towards zero at the beginning of training. This improves the training efficiency and convergence speed of the BP neural network, so we use the Adam algorithm to optimise the BP neural network. The algorithm's update rules are as follows:

	 Update condition:

	 \begin{equation}
	 g=\nabla_{\theta_{k-1}} L(\theta)
	 \end{equation}
	 
	 	 \begin{equation}
	 		 m_{k}=\beta_{1} m_{k-1}+\left(1-\beta_{1}\right) g 
	 	 \end{equation}

	 \begin{equation}
		 		 v_{k}=\beta_{2} m_{k-1}+\left(1-\beta_{2}\right) g \odot g 
		 	 \end{equation}

		 \begin{equation}
			 	 \hat{m}_{k}=\frac{m_{k}}{1-\beta_{1}^{k}}
			 	 \end{equation}
	
		 \begin{equation}
				 		 \hat{v}_{k}=\frac{v_{k}}{1-\beta_{2}^{k}} 
				 	 \end{equation}
	 
	 	 \begin{equation}
	 	 				 		 \theta_{k}=\theta_{k-1}-\frac{\eta}{\sqrt{\hat{v}_{k}}+\varepsilon} \hat{m}_{k}
	 	 				 	 \end{equation}

Among them,  $m_{\mathrm{k}} $ and $ v_{\mathrm{k}} $ are the estimates of the first and second moments of the gradient, respectively, and are the exponential decay rates that control these two moment estimates, which are usually set to 0.9 and  0.999 . $\epsilon$  is a very small number (e.g., 1e-8) to prevent division by zero. k is the current iteration number, which is used for bias correction.
\begin{equation}
\beta_{1}^{k}=\beta_{1} \times \beta_{1} \times \cdots \times \beta_{1} \quad(\text { Multiply by kkk times }) 
\end{equation}
\begin{equation}
\beta_{2}^{k}=\beta_{2} \times \beta_{2} \times \cdots \times \beta_{2} \quad(\text { Multiply by kkk times })
\end{equation}

In the Adam optimization algorithm, $ \beta_{1}^{k} $ and $ \beta_{2}^{k}$  are used for bias correction. This is because at the beginning of the algorithm, $ m_{k} $ and $ v_{k} $ (estimates of the first and second moments of the gradient, respectively) are initialized from 0 , which causes them to be underestimated in the initial stage. This bias is more pronounced when  $\beta_{1} $ and $ \beta_{2} $ are close to 1 . To compensate for this bias in the estimates, the Adam algorithm introduces a bias correction mechanism, which improves the accuracy and stability of training. However, in some cases, Adam may lead to overfitting of the model, resulting in unsatisfactory results.

In this algorithm, the L2 regularization optimization algorithm is used, and the sum of the squares of the weights is added to the loss function. This makes the weights evenly small, avoids some weights being too large while others are too small, improves the generalization ability of the neural network, and prevents overfitting[25-26].
\begin{equation}
L(\omega)=\frac{1}{N} \sum_{i=1}^{N}\left(f\left(x_{i} ; \omega\right)-y_{i}\right)^{2}+\frac{\lambda}{2}\|\omega\|^{2}
\end{equation}

 $ \lambda  $ is the regularization parameter, and  $ \|\omega\|^{2} $ is the square of the Euclidean norm of $ \omega $.
  
The optimisation result is shown in the figure below:
The experimental results show that, as shown in Figure 7, Figure (a) is the prediction result of the test set of the optimized BP neural network, with a prediction accuracy of 96.69$\%$ and a good prediction result. Figure (b) shows that the best training performance was 0.00218 in the sixth round. Compared with the traditional BPNN, it can be seen that the number of iterations has increased by two rounds, but the mse has been effectively reduced. Figure (c) shows the error histogram. It can be seen from the image that the recognition error of the optimized neural network is small, close to the zero error value, and the performance of the optimized BP neural network is better. Figure (d) shows the ROC image. The ROC curve is biased towards the upper left corner, indicating that the Receiver Operating Characteristic curve is better. 

     \begin{figure}[!h]
      \centering
      \includegraphics
        [width=0.9\hsize]
        {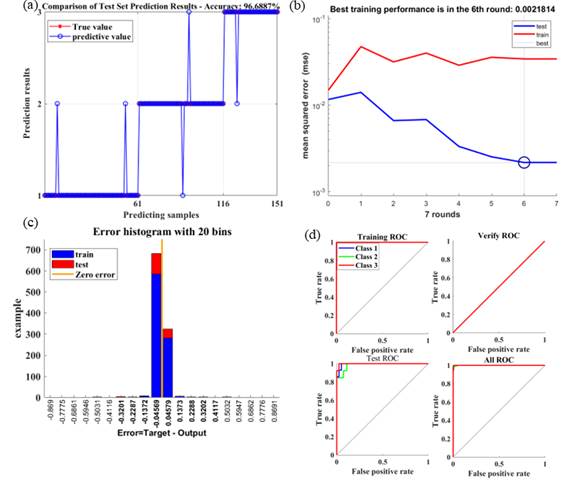}
   
      \caption{Figure 7. Optimised test results (a) The test set flow pattern has a prediction accuracy of 96.69$\%$. (b) The best training performance was 0.00218 in the sixth round. (c) Error histogram, the recognition error of the optimised neural network is small and close to the zero error value. (d) ROC image, the ROC curve is biased towards the upper left corner.
      }
      \label{fig:one-column-figure}
      \end{figure}

By comparing the predicted results of the training set and the test set, we found that the results are consistent with the actual situation. In actual production and experimentation, in a few cases, because the physical parameters such as the apparent velocities of the two phases of the two flow patterns are the same, the void fractions of the two flow patterns are similar, and the types of flow patterns cannot be well distinguished. In this case, the bubble flow with a higher void fraction may be misidentified as a ring flow. This is because in order to better collect data, the fluid is set to be incompressible, and the flow pattern does not change. In actual production, as the height of the vertical rising pipe increases, the pressure gradually decreases, the bubbles gradually become larger, and the void fraction becomes larger, which naturally accompanies the transformation of the flow pattern. Therefore, the bubble flow with a larger flow rate is mistakenly identified as a ring flow. A small amount of ring flow close to the wall is mistakenly identified as bubble flow. This is unavoidable in the collection of actual production parameters. To reduce the measurement of misleading data, the probe structure can be improved and the multi-probe method can be used for flow pattern measurement.

Based on the above experimental principles, we have used an artificial intelligence BP neural network model to predict the gas-liquid two-phase flow pattern in most vertical rising pipes. This method provides a simple and efficient tool for studying and predicting the flow behaviour of gas-liquid two-phase flow, and provides a reliable basis for the design and optimisation of related engineering fields.

\begin{table}[!h]
\centering
\caption{ Flow pattern recognition rates for different classifiers}
\label{tab:my-table}
\begin{tabular}{ll}
\hline
Classifier                & Recognition rate (\%) \\\hline
BPNN                      & 90.73                 \\
Nesterov- BPNN            & 91.39                 \\
Adam-BPNN                 & 92.71                 \\
Regularization- BPNN      & 95.36                 \\
Regularization- Adam-BPNN & 96.68                  \\\hline
\end{tabular}
\end{table}

In addition, 504 sets of data were input into the BPNN, Nesterov-BPNN, Regularization-BPNN, Adam-BPNN, and Regularization-Adam-BPNN classifiers to identify the flow pattern of two-phase flow. Table 1 shows the classification accuracy of flow patterns based on different optimisation algorithms. It can be seen that the regularized Adam-BPNN algorithm has the best recognition rate, up to 96.68$\%$, while the traditional BPNN has the lowest recognition rate of 90.73$\%$. The recognition rate of Nesterov-BPNN is 91.39$\%$, The Nesterov momentum method does not have the same high recognition rate as the Adam optimizer, which shows that our test scenario is more suitable for using the Adam optimizer. The recognition rate of Adam-BPNN is 92.71$\%$, which may be because the Adam optimizer has performed bias correction on the BPNN, which reduces the impact of the initial iterations and improves the recognition rate. However, the Adam optimizer is is more sensitive to parameters. The recognition rate of Regularization-BPNN is 95.36$\%$, which may be because the parameters we set at the beginning were not reasonable, and too many or too few neurons were used in the hidden layer, resulting in overfitting or underfitting of the neural network. After we used neural network regularization, the recognition rate of the neural network improved significantly. To sum up, we use the regularized Adam-BPNN algorithm to recognize two-phase flow patterns in order to improve the recognition rate of flow patterns.
The confusion matrix for the recognition results of all test samples by the different classifiers is shown in Figure 8. A total of 151 test samples were recognised in this study. Due to the random seed we set, the number of each flow pattern recognised in each test was different. As can be seen in Figure 8, of the five classifiers, Regularization-Adam-BPNN has a high recognition rate for each flow pattern, with only six flow patterns being misidentified. Nesterov-BPNN, Adam-BPNN and Regularization-BPNN all have lower flow pattern recognition performance than Regularization-Adam-BPNN, with 13, 11 and 7 flow patterns being misidentified respectively. The traditional BPNN had the worst recognition performance, with 14 flow types being misidentified.
 
    \begin{figure}[!h]
       \centering
       \includegraphics
         [width=0.95\hsize]
         {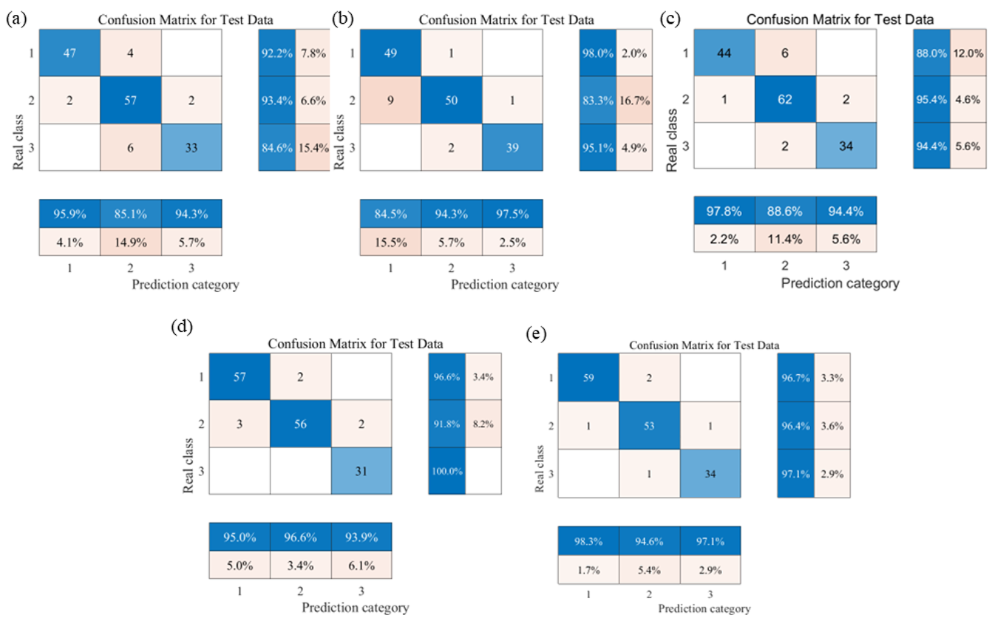}
         
       \caption{Figure 8. Confusion matrix of different classifiers. The left side of the picture shows the real class, and the bottom side shows the predicted class. 1 represents bubble flow, 2 represents elastic flow, and 3 represents annular flow. Figure (a) shows the confusion matrix of the BPNN, where 4 bubble flows were identified as elastic flows. 2 elastic flows were identified as bubble flows, and 2 elastic flows were identified as annular flows. Six annular flows were identified as bullet-shaped flows. Figure (b) shows the confusion matrix of the Nesterov-BPNN, with one bubble flow identified as a bullet-shaped flow. Nine bullet-shaped flows were identified as bubble flows, and one bullet-shaped flow was identified as an annular flow. Two annular flows were identified as bullet-shaped flows. Figure (c) shows the confusion matrix of the Adam-BPNN, with 6 bubble-like flows being identified as bullet-like flows. 1 bullet-like flow was identified as a bubble-like flow, and 2 were identified as ring-like flows. 2 ring-like flows were identified as bullet-like flows. Figure (d) shows the confusion matrix of the Regularization-BPNN, with 2 bubble flows being identified as bullet flows. 3 bullet flows are identified as bubble flows, and 2 are identified as ring flows. All the bubble-like flows are recognised. Figure (e) shows the confusion matrix of the Regularization-Adam-BPNN, with two bubble-like flows being recognised as bubble-like flows. One bubble-like flow is recognised as a bubble-like flow, one bubble-like flow is recognised as a ring-like flow, and one ring-like flow is recognised as a bubble-like flow.
       }
       \label{fig:one-column-figure}
       \end{figure}

\section{Conclusion}
Due to the equipment level and experimental conditions of the two-phase flow dynamic experimental platform, this paper mainly uses Fluent software for simulation. The void fraction data of three flow patterns: bubble flow, elastic flow and annular flow with different two-phase flow velocities are obtained. Then, based on the Adam-optimised BP neural network, the neural network is regularised to achieve the identification of the flow pattern. A study of flow pattern recognition was carried out for bubble flow, elastic flow and annular flow in gas-liquid two-phase flow, and a flow pattern recognition rate of 96.68$\%$ was successfully achieved. This provides a valuable reference for flow pattern recognition technology in actual production. However, this is still not enough. In this regard, we can use multi-probe detection in actual production to improve the flow pattern recognition rate. Moreover, simulation research itself has certain limitations, and the actual production environment is complex and changeable, and may be a mixture of multiple flow patterns. Therefore, in future work, we will use the multi-probe method to further carry out experimental research and collect data on gas-liquid two-phase flow in industrial sites to compensate for this deficiency. Improving the probe structure, without affecting the original flow of the two-phase flow, determines the flow pattern in the well, visually presenting the situation in the well in front of people, solving practical production problems, and promoting the development

\end{document}